\documentclass[11pt,a4paper]{article}
\usepackage{jcappub-modified}
\usepackage{bm,verbatim}

\def\la{~\mbox{\raisebox{-.6ex}{$\stackrel{<}{\sim}$}}~}
\def\ga{~\mbox{\raisebox{-.6ex}{$\stackrel{>}{\sim}$}}~}

\makeindex

\title{Blue Tensor Spectrum from Particle Production during Inflation}

\author[a]{Shinji Mukohyama,}
\author[a]{Ryo Namba,}
\author[b]{Marco Peloso}
\author[c,d]{and Gary Shiu}

\affiliation[a]{Kavli Institute for the Physics and Mathematics of the Universe (WPI), \\
Todai Institutes for Advanced Study, University of Tokyo, Kashiwa, Chiba 277-8583, Japan}
\affiliation[b]{School of Physics and Astronomy, University of Minnesota, Minneapolis, MN 55455, USA}
\affiliation[c]{Department of Physics, University of Wisconsin, Madison, WI 53706, USA}
\affiliation[d]{Center for Fundamental Physics and Institute for Advanced Study, \\
Hong Kong University of Science and Technology, Hong Kong}

\emailAdd{shinji.mukohyama@ipmu.jp}
\emailAdd{ryo.namba@ipmu.jp}
\emailAdd{peloso@physics.umn.edu}
\emailAdd{shiu@physics.wisc.edu}

\abstract{
We discuss a mechanism of particle production during inflation that can result in a blue gravitational wave (GW) spectrum, compatible with the BICEP2 result and with the $r < 0.11$ limit on the tensor-to-scalar ratio at the Planck pivot scale. The mechanism is based on the production of vector quanta from a rolling pseudo-scalar field. Both the vector and the pseudo-scalar are only gravitationally coupled to the inflaton, to keep the production of inflaton quanta at an unobservable level (the overproduction of non-gaussian scalar perturbations is a generic difficulty for  mechanisms that aim to generate a visible GW signal from particle production during inflation). This mechanism can produce a detectable amount of GWs for any inflationary energy scale. The produced GWs are chiral and non-gaussian; both these aspects  can be tested with large-scale polarization data (starting from Planck). 
We study how to reconstruct the pseudo-scalar potential from the GW spectrum. 
}

\begin{document}

\maketitle

\section{Introduction}

Cosmic inflation in the early universe is arguably the most promising candidate
for the origin of primordial fluctuations, from which the rich
structures of our universe such as galaxies and clusters of galaxies
emerged due to gravitational instability.

One of the most important and robust predictions of inflationary
cosmology is that it generates quantum fluctuations of the graviton, a
spin-$2$ degree of freedom that mediates gravity, leading to a
tensor-type perturbation, a.k.a. gravitational waves (GWs). In general relativity, the power spectrum of
tensor perturbation from vacuum fluctuations during inflation is given
by  
\begin{equation}
 P_{\rm GW}(k) =  
  \left.\frac{2H^2}{\pi^2M_{\rm Pl}^2}\right|_{k=aH} \; .
  \label{eqn:tensor-standard}
\end{equation} 
This depends only on the Hubble expansion rate $H$ and, for this reason, the
tensor power spectrum is usually considered as a direct probe of the
scale of inflation. For the same reason, it is commonly believed that
the tensor spectrum from inflation is always red: the Hubble expansion
rate gradually decreases during inflation and thus modes with shorter
wavelengths have lower amplitudes than those with longer
wavelengths.

With this interpretation of tensor spectrum, detection of primordial
GWs within the current observational reach would suggest
that cosmic inflation actually occured in our universe at rather high
energy scale, potentially ruling out many low-scale models of
inflation. Moreover, it would provide a strong evidence that the
graviton indeed exists and follows the laws of quantum mechasnics. For
this reason, the recent detection of B-mode polarization by BICEP2 
collaboration \cite{Ade:2014xna}, if confirmed, would be a strong motivation for further
studies of inflationary scenarios in the context of quantum gravity such
as string theory~\footnote{Inflation models in string theory
 are so far mostly low scale ones \cite{Baumann:2014nda}.
An exception is axion monodromy inflation
\cite{Silverstein:2008sg,McAllister:2008hb,Marchesano:2014mla}. The new class of monodromy inflation models with F-term potentials  \cite{Marchesano:2014mla} naturally evades the eta problem
and several microphysical constraints. See also \cite{Kaloper:2008fb,Kaloper:2011jz,Kaloper:2014zba} for field theoretical descriptions and \cite{other_monodromy} for some recent  model building efforts.}. However,
these considerations are only relevant if the formula
(\ref{eqn:tensor-standard}) is valid. It is thus important to
identify the regime of validity of this standard result, and explore novel ways to evade it.

The purpose of the present paper is to point out a possibility to enhance the tensor power spectrum 
from the standard formula (\ref{eqn:tensor-standard}) in a scale-dependent way. The standard result is the power spectrum of tensor modes obeying the free field equation
\begin{equation}
\left[ \partial_\tau^2 + k^2 - \frac{a''}{a} \right] \left( a \, \delta g_{ij} \right) = S_{ij} \;\;,\;\; S_{ij} = 0 \,, 
\label{vacuum-eq}
\end{equation} 
where $a$ is the scale factor of the universe,  $\tau$ is the conformal time, $k$ is the comoving  momentum, and $\delta g_{ij}$ are the tensor perturbations of the metric. Due to the expansion of the universe, the graviton wave function is ``dragged out'' from the empty vacuum state to the power  (\ref{eqn:tensor-standard}). Strictly speaking, the right-hand side of eq.~(\ref{vacuum-eq}) never vanishes, as the graviton is nonlinearly coupled to itself and to any other field. However the coupling is of gravitational strength, and the source term $S_{ij}$ is second order in perturbation theory, and for these reasons $S_{ij}$ is typically negligible. Recently, the works \cite{Cook:2011hg,Senatore:2011sp,Barnaby:2012xt,Carney:2012pk} have investigated the possibility that a substantial particle production takes place during inflation, with the aim of  obtaining a source $S_{ij}$ for the GWs that can dominate the observed tensor modes from inflation.  The mechanism mostly studied in 
\cite{Cook:2011hg,Senatore:2011sp,Carney:2012pk} consists in introducing a field $\chi$ that acquires a mass from its interaction with the inflaton, $M_\chi = M \left( \phi \right)$. This mass is arranged such that $M=0$ for some given value $\phi = \phi_*$ acquired by the inflaton during inflation. As $\phi$ crosses this value, a burst of quanta of $\chi$ is produced nonperturbatively \cite{Chung:1999ve}, which are the GW source $S_{ij}$. The minimal implementation of this mechanism, however, gives rise to a negligible GW production \cite{Barnaby:2012xt}: as the inflaton moves past $\phi_*$, the quanta of $\chi$ quickly become very massive and non-relativistic. This strongly suppresses their quadrupole moment (the source of GWs). On the other hand, the same quanta also source highly non-gaussian inflaton perturbations \cite{Barnaby:2012xt}. This production is not suppressed by the small quadrupole moment of the non-relativistic quanta, and thus the non-observation of this non-gaussian signal puts a phenomenological constraint on this mechanism.~\footnote{This is true even if the scalar field that controls the mass of $\chi$ is not the inflaton; inflaton quanta are at least gravitationally coupled to $\chi$, and they are produced in a much greater amount than GWs  \cite{Barnaby:2012xt}.} 

One may imagine modifying this implementation in ways that can enhance the GW signal. For example, 
ref.~\cite{Senatore:2011sp} mentioned the possibility that the particles $\chi$ decay, and that their decay products source GWs. However, it is hard to reconcile this with the non-gaussianity bounds   \cite{Barnaby:2012xt}. 

A more promising way to enhance the ratio between the sourced tensor and the sourced scalar modes is to have a copious production of quanta that remain relativistic. A vector field coupled to a rolling pseudo-scalar can experience a significant tachyonic growth during inflation \cite{Anber:2009ua}. If the rolling pseudo-scalar is the inflaton,  also this mechanism results in an unobservable GW signal, due to the limits imposed by the non-gaussianity of the inflaton perturbations \cite{Barnaby:2010vf,Barnaby:2011vw}. On the other hand, if the rolling pseudo-scalar is a different field $\psi$, and if the inflaton is coupled only gravitationally to the vector field, then the vector quanta produce a greater amount of GWs than of inflaton perturbations, and the sourced GW signal can dominate over the vacuum one, while the sourced inflaton quanta are negligible  \cite{Barnaby:2012xt}.%
\footnote{We are assuming that the energy density $\rho_\psi$ of the pseudo-scalar becomes sufficiently small by the end of inflation, so that  the pseudo-scalar does not contribute to the late time observable curvature perturbation $\zeta_{\rm obs.}$. Strictly speaking, for $\rho_\psi \neq 0$ at reheating, the pseudo-scalar contributes to the observable curvature perturbations. As $\psi$ is coupled more strongly than gravitationally to the vector field $A$, this would generate a component of $\zeta_{\rm obs.}$ that is sourced more strongly than gravitationally by $\delta A$. This component is proportional to $\rho_\psi$ at reheating, so it does not impact observations if $\rho_\psi$ is much smaller than $\rho_\phi$ at reheating. It is very easy to achieve this. For example, if the pseudo-scalar reaches the minimum of its potential and becomes massive during inflation (say a few e-folds after the CMB modes have left the horizon), then its energy density (or the one of its decay products) redshifts away to a completely negligible value as compared to the energy density of the slow-rolling inflaton.}
To our knowledge, this is the only model of particle production during inflation which has been proven to (i) produce a significant amount of GWs, and (ii) avoid too strong  non-gaussianity of the scalar perturbations, and for this reason we study this model in this work. 

On the other hand, the tensor modes in this model are highly non-gaussian, and they can generate an interesting equilateral non-gaussianity in the CMB temperature perturbations and polarization \cite{Cook:2013xea,Shiraishi:2013kxa}. We show that, based on the results of these works, the current Planck limit on equilateral non-gaussianity does not yet constrain the mechanism. However, an interesting non-gaussian signature may emerge from  the Planck polarization data and from future CMB experiments.

A second very interesting signature of this model is a breaking of parity in the GW signals, due to the fact that only the GWs of one given helicity are significantly produced.%
\footnote{
A chiral GW background from inflation was first considered in \cite{Lue:1998mq}, where parity violation arises from a gravitational Chern-Simons term coupled to inflaton.
}
This can lead to  signatures in the TB and EB correlations of the CMB anisotropy. Refs.~\cite{Saito:2007kt,Gluscevic:2010vv,Ferte:2014gja} studied the detectability of this signal as a function of  $r$ and of the parity violation, showing that most of the contribution to this detection comes from the very largest scales, $\ell \lesssim 10$ (therefore this effect does not impact the BICEP2 measurements or forthcoming missions with a small sky coverage \cite{Ferte:2014gja}). If the signal measured by BICEP2 is cosmological, and it is dominated by the GWs produced by our mechanism, the parity violation may be already detectable at the $1 \sigma$ level in the Planck polarization data \cite{Gluscevic:2010vv}, and a more significant detection should be expected in future full sky or nearly full sky experiments \cite{Ferte:2014gja}. 

In this model, GWs are produced continuously during inflation, with an amplitude exponentially sensitive to the speed of the pseduo-scalar field $\psi$, which is not the inflaton. Therefore, the  resulting tensor spectrum can be either blue or
red, or can even have bumps, depending on the pseudo-scalar potential. As we will see, these features allow to reconcile the possible tension between Planck/WMAP and BICEP2. The Planck temperature data, supplemented with priors from the WMAP polarization measurements, provide the limit $r\la 0.11$ on the tensor-to-scalar ratio at the $k_0 = 0.002 \, {\rm Mpc}^{-1}$ pivot scale \cite{Ade:2013zuv}. This is in contrast with the BICEP2 detection $r =0.20_{-0.05}^{+0.07}$ (without polarized dust removal) or $r =0.16_{-0.05}^{+0.06}$ (with the dust removal method that they deem most reliable) \cite{Ade:2014xna} at somewhat smaller angular scales (for definiteness, we take  $k_0 = 0.0057 \, {\rm Mpc}^{-1}$, corresponding to $\ell \approx 80$). A solution to this possible tension may be a blue tilt in the tensor spectrum \cite{Gerbino:2014eqa}, which would then result in a running of the tensor-to-scalar-ratio,  ${\cal T} _r \equiv \frac{d \ln r}{d \ln k}=n_T-(n_s-1)$ \cite{Gong:2007ha}, where $n_s$ and $n_T$ are the scalar and tensor spectral indices, respectively. While standard single-field slow-roll inflation satisfies $n_T=-r/8$, a significant blue tensor tilt can be attained -- besides from the particle production mechanism that we study here -- with generalized slow-roll \cite{Gong:2014qga}, non-Bunch Davies initial states \cite{Ashoorioon:2014nta} (extending \cite{Ashoorioon:2013eia}),
sourcing from a spectator field with small sound speed \cite{Biagetti:2013kwa}, other non-standard inflationary scenarios \cite{Brandenberger:2014faa,Wang:2014kqa,Mohanty:2014kwa}, and some non-inflationary mechanisms \cite{Biswas:2014kva}. Other ideas to reconcile this possible tension include invoking a negative running of the scalar spectral index \cite{Ade:2014xna,Nakayama:2014koa,Ma:2014vua,Germani:2014hqa,McDonald:2014kia,Sloth:2014sga,Hu:2014aua,Kinney:2014jya}, suppression of  the scalar power at large scales \cite{Contaldi:2014zua,Miranda:2014wga,Abazajian:2014tqa,Hazra:2014aea,Bousso:2014jca,Firouzjahi:2014fda},  anti-correlated isocurvature perturbations \cite{Kawasaki:2014lqa,Kawasaki:2014fwa,Bastero-Gil:2014oga},  a space-dependent $r$ \cite{Chluba:2014uba},  cosmological birefringence \cite{Lee:2014rpa},  dark radiation \cite{Giusarma:2014zza,Ishida:2014zya,Xu:2014laa},  primordial magnetic fields \cite{Bonvin:2014xia}, a pre-inflationary bounce \cite{Xia:2014tda}, and sterile neutrinos \cite{Zhang:2014dxk,Dvorkin:2014lea,Anchordoqui:2014dpa,Zhang:2014nta,Leistedt:2014sia}.  

We note that the relation $n_T = - r/8$ is also violated in warm inflation \cite{Berera:1995ie}, as the tensor-to-scalar ratio is in this case reduced compared to the standard case due to the dissipative effects and the much greater production of scalar with respect to tensor perturbations due to the coupled inflaton \cite{BasteroGil:2009ec}.

In the model we study in this paper, the tensor spectrum contains useful information about the properties of the pseudo-scalar field. In particular we show that one can in principle reconstruct the potential of the pseudo-scalar field from the tensor spectrum, provided that the measured tensor spectrum in our universe is higher than that given by the formula (\ref{eqn:tensor-standard}). As examples, we study two different cases, one characterized by a continuously growing particle production, and consequently continuous growth in the tensor power (under the assumption that this continuous growth takes place for the scales relevant for the CMB observations), and one by a sudden but momentary acceleration of the pseudo-scalar motion, resulting in a localized burst of particle production and a localized growth of $r$.

This work is organized as follows. In Section \ref{model} we summarize
the model, first introduced in ref.~\cite{Barnaby:2012xt}. In Section
\ref{gaugeprod} we discuss the gauge field production in this model, and
the several consequent observational implications of the resulting GW
background for  the CMB observations. In section \ref{blue} we emphasize
that the GW can be blue in this model. In section
\ref{sec:reconstruction} we show how, at least in principle, the
potential of the pseudo-scalar field can be reconstructed from the GW
spectrum. Finally, in Section \ref{sec:conclusions} we conclude the
paper.

\section{Model description}
\label{model}

We consider the model introduced in \cite{Barnaby:2012xt}, in which an inflaton $\varphi$ and a pseudo-scalar field $\psi$ are minimally coupled to the Einstein gravity. The pseudo-scalar field also couples to a $U(1)$ gauge field in a way consistent with symmetries. Assuming canonical kinetic terms for $\varphi$ and $\psi$, the action is 
\begin{equation}
 S = \int d^4x\sqrt{-g}
  \left[ \frac{M_{\rm Pl}^2}{2}R
   - \frac{1}{2}(\partial\varphi)^2 - V(\varphi) 
   - \frac{1}{2}(\partial\psi)^2 - U(\psi)
   - \frac{1}{4}F^2 - \frac{\psi}{4f}F\tilde{F}
       \right] \; ,
\label{action}
\end{equation}
where $F$ and $\tilde{F}^{\mu \nu} \equiv \frac{1}{2} \epsilon^{\mu \nu \alpha \beta} F_{\alpha \beta}$ are the field-strength tensor of the gauge field and its dual, respectively, and $f$ is a pseudo-scalar decay constant \footnote{Note our normalization for $f$ is $4 \pi^2$ times larger than the conventional normalization used e.g. in \cite{Choi:1985je,Banks:2003sx,Svrcek:2006yi}.}. We assume that $\varphi$ and $\psi$ take homogeneous vacuum expectation values (vev), $\bar{\varphi} (t)$ and $\bar{\psi} (t)$, respectively, and that the background spacetime is of the flat FLRW form with the metric
\begin{equation}
 ds^2 = -dt^2 + a(t)^2\, (dx^2+dy^2+dz^2) \; ,
\end{equation} 
where $a(t)$ is the scale factor. We assume that the gauge field carries no vev.

The equations of motion for $\bar \varphi$ and $\bar \psi$ are
\begin{eqnarray}
 \ddot{\bar \varphi} + 3H\dot{\bar \varphi} + V'(\bar \varphi)  & = & 0 \; , \\
  \ddot{\bar \psi} + 3H\dot{\bar \psi} + U'(\bar \psi)  & = & 0 \; .
\label{eqn:EOMpsi}
\end{eqnarray}
where $H\equiv \dot{a}/a$ is the Hubble parameter, and an overdot and a prime denote the derivatives with respect to $t$ and to the argument, respectively. We assume that the contribution of the pseudo-scalar $\psi$ to the background evolution is negligible compared to that of the inflaton $\varphi$, requiring
\begin{equation}
 |U| \ll V\,, \quad 
  \dot{\bar \psi}^2 \ll \dot{\bar \varphi}^2 \; .
\end{equation}
Under this condition, the Einstein equations for the background are approximated as
\begin{eqnarray}
 3M_{\rm Pl}^2 H^2 & \simeq & 
  \frac{1}{2}\dot{\bar \varphi}^2 + V(\bar \varphi) 
  \simeq V (\bar \varphi) \; , 
  \label{eqn:Friedmann} \\
 2M_{\rm Pl}^2\dot{H} & \simeq & -\dot{\bar \varphi}^2 \; ,
  \label{eqn:dynamical}
\end{eqnarray} 
where the last approximate equality in the first line assumes the slow roll of the inflaton.

\section{Gauge-field production and its effects}
\label{gaugeprod}

The coupling of the gauge field $A_\mu$ to the time-dependent pseudo-scalar vev $\bar \psi$ leads to a copious production of the gauge quanta. The linearized equation of motion for the Fourier modes of the gauge field is \cite{Anber:2009ua}
\begin{equation}
\left[ \partial_\tau^2 + k^2 \pm \frac{2 k \xi}{\tau} \right] 
 A_\pm (\tau , k) \simeq 0\,, \quad 
 \xi \equiv \frac{\dot{\bar \psi}}{2 H f}\; ,
\label{eom-gauge}
\end{equation}
where $\tau$ is the conformal time $d \tau = dt / a$, and $\pm$ corresponds to the circular polarization states of the gauge field. We assume that the motion of the pseudo-scalar vev is over-damped and thus treat $\xi$ as nearly  constant.
This condition reads~\footnote{Violation of the condition (\ref{def-deltaxi}) does not necessarily mean that the
particle production and the subsequent generation of GWs
are inefficient. In fact it would be interesting to extend this computation outside the regime (\ref{def-deltaxi}). We hope to come back to this issue in a future publication.}
\begin{equation}
 |\delta_{\xi}| \ll 1\,, \quad 
  \delta_{\xi} \equiv \frac{\dot{\xi}}{\xi H}\; .
\label{def-deltaxi}
\end{equation}
Also we assume that $\dot{\bar \psi} > 0$ so that $\xi$ is positive, and
the ``$+$'' state experiences a tachyonic growth near the horizon
crossing while the ``$-$'' state stays in the vacuum (the opposite happens if  
 $\dot{\bar \psi}  $ is negative, and our study can be immediately applied also to this case). 
 The production of the ``$+$'' helicity
modes can be well quantified by the approximate solution
\cite{Anber:2009ua} 
\begin{equation}
A_+ \simeq \left( \frac{- \tau}{8 \xi k} \right)^{1/4} \, {\rm e}^{\pi \xi - 2 \sqrt{ -2 \xi k \tau} } \; , \quad \partial_\tau A_+ \simeq \sqrt{\frac{2 \xi k}{-\tau}} \, A_+ \; .
\label{gauge-sol}
\end{equation}
We restrict our consideration to $\xi \gtrsim 1$, for which each mode experiences a significant exponential growth ${\rm e}^{\pi \xi} \gg 1$.

In order to avoid a significant backreaction of the produced gauge quanta to the background dynamics, we must simultaneously require (i) that the energy density of the produced gauge field is smaller than the kinetic energy of $\bar \psi$ and (ii) that the backreaction to the equation of motion for $\bar{\psi}$, eq.~(\ref{eqn:EOMpsi}), is negligible. It has turned out that (i) is more stringent a constraint than (ii) and requires \cite{Barnaby:2012xt}
\begin{equation}\label{backreaction}
\frac{{\rm e}^{\pi \xi}}{\xi^{5/2}} \ll \frac{13.5}{\sqrt{ \epsilon \, {\cal P}}} \, \frac{f}{M_p} \; ,
\end{equation}
where $\epsilon \equiv - \dot{H} / H^2$ is the slow-roll parameter, and ${\cal P} \equiv H^2 / \left( 8 \pi^2 \epsilon M_{\rm Pl}^2 \right)$ corresponds to the result for $P_\zeta$ without source. Under this condition, the produced gauge field does not alter the background dynamics.

The gauge field is gravitationally coupled both to the inflaton
perturbations and to the GWs, as can be seen from (\ref{action}). Thus the
produced gauge quanta (\ref{gauge-sol}) in turn source the scalar and
tensor perturbations through the Einstein equations. This sourcing effect
induces a contribution to their spectra that is uncorrelated to that from
the standard vacuum fluctuations. The total spectrum of the scalar
perturbation was found in \cite{Barnaby:2012xt} to be  
\begin{equation}
P_\zeta \simeq {\cal P} \left( 1 + 2.5 \cdot 10^{-6} \, \epsilon^2 \, {\cal P} \, \frac{{\rm e}^{4 \pi \xi}}{\xi^6} \right) \; .
\label{Pzeta}
\end{equation}
The second term in the parenthesis is the contribution sourced by the gauge field while the first is that from the vacuum fluctuations.
On the other hand, the GW power spectrum was computed in \cite{Barnaby:2010vf, Sorbo:2011rz},
\begin{equation}
P_{{\rm GW}} \simeq 16 \, \epsilon \, {\cal P} \left( 1 + 3.4 \cdot 10^{-5} \, \epsilon \, {\cal P} \, \frac{{\rm e}^{4 \pi \xi}}{\xi^6} \right) \; ,
\label{PGW}
\end{equation}
where again the first and second terms in the parenthesis correspond to the contributions from the vacuum fluctuations and from the source, respectively.

\begin{figure}
\centering
\includegraphics[width=0.7\textwidth]{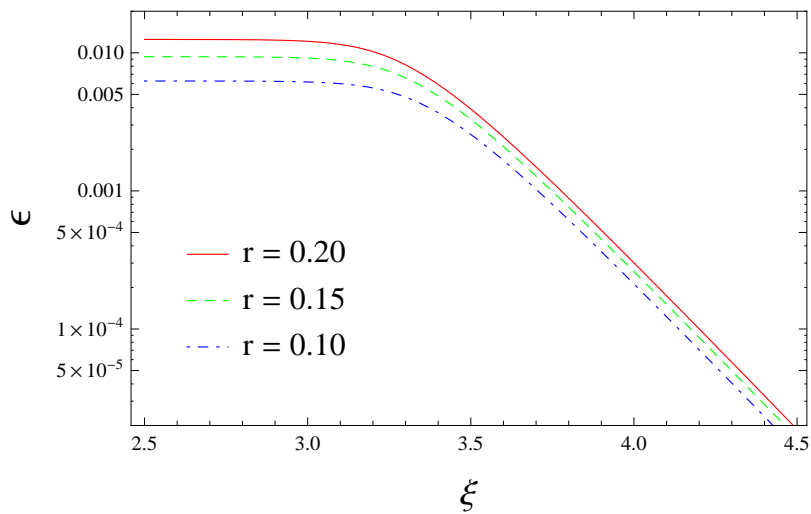}
\caption{Slow roll parameter $\epsilon$ vs particle production parameter $\xi$, when $P_\zeta$ is fixed to the measured amplitude, and for three different values of $r$. At the smallest $\xi$ shown, particle production is negligible, and $\epsilon = r/16$. As $\xi$ increases the sourced GWs become dominant, and $\epsilon$ needs to decrease to keep $r$ at the given value. \label{fig:eps-xi}}
\end{figure}

Combining (\ref{Pzeta}) and (\ref{PGW}) we obtain 
\begin{equation}
 P_\zeta \simeq {\cal P} \, \frac{1-0.0735 \epsilon}{1- 0.0046 r} \;\;, 
 \label{eqn:P-Pzeta}
\end{equation} 
showing that the sourced scalar perturbations, $P_{\zeta}-{\cal P}$, are
much smaller than the vacuum ones, ${\cal P}$, in this model. On the
contrary, the sourced GWs  dominate over the vacuum ones for $\xi \ga
3.4$ when $r \simeq 0.2$. In Figure \ref{fig:eps-xi}  we plot $\epsilon$ as a function of $\xi$, once the scalar spectrum is normalized to the observed value,  and once the tensor-to-scalar ratio is fixed to  $r=0.2$ (red solid curve), $r=0.15$ (green dashed curve), or $r=0.1$ (blue dot-dashed curve). The vacuum modes dominate at the
smaller values of $\xi$ shown in the plot, and the standard slow roll
result $\epsilon = r/16$ is obtained there. As $\xi$ increases, the
sourced GW signal dominates, and $\epsilon$ needs to decrease
exponentially. 

The scalar and tensor modes sourced by the vector field are highly non-gaussian. As studied in \cite{Barnaby:2012xt} 
the sourced scalar modes are so small in comparison to the vacuum ones that they do not lead to any observable non-gaussianity. On the other hand, as noted in \cite{Cook:2013xea}, the sourced tensor modes could leave a sizable non-gaussianity of nearly equilateral shape on the CMB anisotropies and polarization. The amount of non-gaussianity  is controlled by the parameter $X$ and estimated as \cite{Cook:2013xea}
\begin{eqnarray}
f_{\rm NL} & \simeq & 1.1 \cdot 10^{-14} \, X^3 \nonumber\\
X & \equiv & \epsilon \, \frac{{\rm e}^{2 \pi \xi}}{\xi^3} \;\;. 
\label{X-def}
\end{eqnarray} 

\begin{figure}
 \begin{center}
\includegraphics[width=0.7\textwidth,angle=0]{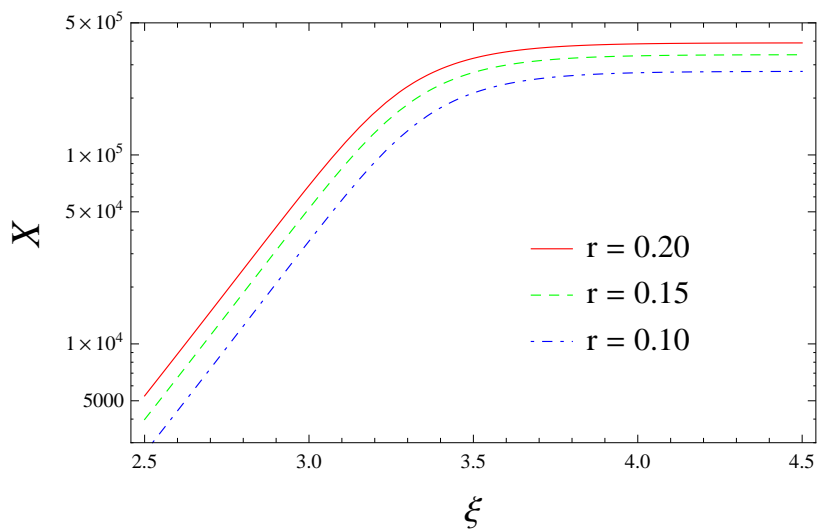}
\caption{Parameter $X$, defined in (\ref{X-def}), controlling the amount of non-gaussianity as a function of the  particle production parameter $\xi$, when $P_\zeta$ is fixed to the measured amplitude, and for three different values of $r$. See the main text for discussion.}
\label{fig:X-xi}
  \end{center}
  \end{figure}

In Figure \ref{fig:X-xi} we plot $X$ as a function of $\xi$, once the scalar spectrum is normalized to the observed value,  and once the tensor-to-scalar ratio is fixed to  $r=0.2$ (red solid curve), $r=0.15$ (green dashed curve), or $r=0.1$ (blue dot-dashed curve). 
The parameter $X$ grows with the amount of sourced GWs, and it saturates to 
\begin{equation}
X \simeq 43 \, \sqrt{\frac{r}{P_\zeta}} \simeq 3.5 \cdot 10^5 \sqrt{\frac{r}{0.15}} \;. 
\label{X-saturation}
\end{equation} 

Using (\ref{X-def}), we see that the $2 \sigma$ Planck temperature limit $f_{\rm NL} < 150$ \cite{Ade:2013ydc} would enforce $X \lesssim 2 \cdot 10^5$ \cite{Cook:2013xea}, if the Planck limit could be taken at face value. However, as also mentioned in \cite{Cook:2013xea}, we cannot take the Planck limit at face value in this case, and this estimate is likely too strong, as the contribution of the tensor modes to the temperature anisotropies scales with $\ell$ differently than the scalar contribution (which the Planck limit assumes) and, in particular, it becomes negligible at $\ell \ga 100$. This is  visible in Figure 1 of \cite{Shiraishi:2013kxa}, which also studied non-gaussianity in this model. According to  \cite{Shiraishi:2013kxa}, a dedicated analysis with the Planck temperature data can detect $X\simeq 5 \cdot 10^5$ at $1 \sigma$. This is above the values obtained in the model (see Figure  \ref{fig:X-xi}). The inclusion of E-mode polarization data can improve the $1 \sigma$ limit to $\simeq 3.8 \cdot 10^5$ and to $\simeq 2.9 \cdot 10^5$ under the Planck and PRISM experiments, respectively. Adding the B mode polarization may allow to probe all the range of $X$ in which the source signal dominates, provided the instrumental error is sufficiently small  \cite{Shiraishi:2013kxa}.

Another interesting feature of the model is that the vev of the
pseudo-scalar breaks parity in the gauge sector. As described right
above (\ref{gauge-sol}), only one of the polarization states of the
gauge field (we take the ``$+$'' state here) is produced.  The
sourcing effect can be understood as the interaction 
$A_+ + A_+ \rightarrow h_R$, with only the right-handed state of the GWs being 
efficiently generated from the gauge sourcing. The contribution from the
vacuum fluctuations of course does not discriminate between the two
helical states. We can quantify the level of chirality using a parameter 
\cite{Saito:2007kt,Barnaby:2012xt} 
\begin{equation}
\Delta \chi \equiv \frac{P_{\rm GW}^R - P_{\rm GW}^L}{P_{\rm GW}^R + P_{\rm GW}^L} \simeq
\frac{3.4 \cdot 10^{-5} \, \epsilon \, {\cal P} \, \frac{{\rm e}^{4 \pi \xi}}{\xi^6}}{1 + 3.4 \cdot 10^{-5} \, \epsilon \, {\cal P} \, \frac{{\rm e}^{4 \pi \xi}}{\xi^6}} \; ,
\label{dchi}
\end{equation}
where $P_{\rm GW}^{R/L}$ is the power spectrum of each state.

\begin{figure}
 \begin{center}
\includegraphics[width=0.7\textwidth,angle=0]{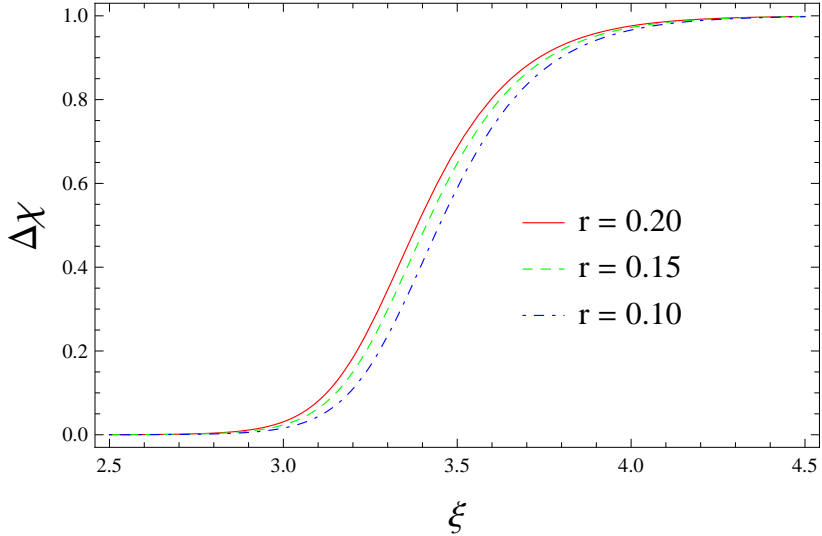}
\caption{Chirality of the observed GWs $\Delta \chi$, defined in (\ref{dchi}), as a function of the  particle production parameter $\xi$, when $P_\zeta$ is fixed to the measured amplitude, and for three different values of $r$. The value of $\Delta \chi$ interpolates from $0$ at small $\xi$ (negligible sourced GWs) to $1$ at large $\xi$ (dominant sourced GWs).}
\label{fig:dc-xi}
  \end{center}
  \end{figure}

In Figure \ref{fig:dc-xi} we show the amount of chirality as a function
of $\xi$, for the same choice of parameters as the previous two
figures. The possible detection of this parity violation in the tensor
sector has been studied in \cite{Saito:2007kt,Gluscevic:2010vv,Ferte:2014gja}. In
particular, ref.~\cite{Gluscevic:2010vv} presented the forecast for
detecting this signal with Planck, Spider, CMBPol, and a hypothetical
cosmic variance limited experiment. Such a violation may be observed at
the $1\sigma$ level in Planck and Spider, provided the sourced signal is
dominant, while a more significant detection can be expected from CMBPol and
from a cosmic variance limited experiment. See Figure 4 of
\cite{Barnaby:2012xt} for the $1 \sigma$ contours in the $\Delta \chi-r$
plane. Ref.~\cite{Ferte:2014gja} further demonstrated that a detection with at least $2 \sigma$ is possible for maximally chiral GWs of $r \gtrsim 0.05$, using the EPIC-2m specifications. As mentioned in \cite{Gluscevic:2010vv,Ferte:2014gja}, on the other hand, the forecasted constraints (or signals) mostly come from low multipoles  $\ell \la 10$, so that we do not expect that the BICEP2 signal provide constraints on this parity violation (and indeed their jackknife $\langle T B \rangle$ and $\langle E B \rangle$ signals appear to be consistent with zero).

\section{Blue tensor spectrum}
\label{blue}

In the previous sections, the parameter $\xi$ has been treated as a constant to obtain (\ref{Pzeta}) and (\ref{PGW}); however, in reality it is a reasonable assumption that $\xi$ is an increasing function of time for some duration during inflation. As long as $\vert \delta_\xi \vert \ll 1$, where $\delta_\xi$ was defined in (\ref{def-deltaxi}), the rate of change is small enough for (\ref{Pzeta}) and (\ref{PGW}) to be valid, up to undetectable corrections.  However, as we now discuss, the increase of $\xi$ can result in a growth of the tensor power at increasingly smaller scales. 

We can  readily compute the tensor spectral tilt from (\ref{PGW}). Using
\begin{equation}
\frac{d \ln H}{d \ln k} \simeq - \epsilon \; , \quad \frac{d \ln \xi}{d \ln k} \simeq \delta_\xi \; ,
\label{xi-tilt}
\end{equation}
up to the first order in slow roll, one finds the tensor spectral tilt as
\begin{equation}
n_T \equiv \frac{d \ln P_{\rm GW}}{d \ln k} \simeq \left( 4 \pi \xi - 6 \right) \delta_\xi - 4 \, \epsilon \; .
\label{tensor-tilt}
\end{equation}
Note that this relation is valid when the sourced contribution dominates
$P_{\rm GW}$. Recall that in order for our calculation to be valid,
$|\delta_\xi| \ll 1$ and $\epsilon\ll 1$ are required. For modestly
large values of $\xi$, there is room to have 
$0 < n_T \lesssim {\cal O} (1)$ while these requirements are satisfied.

\section{From tensor spectrum to axion potential}
\label{sec:reconstruction}

In this section we explicitly show how within this mechanism the potential of the pseudo-scalar $\psi$ can be reconstructed from the GW spectrum. The tensor spectrum is given by (\ref{PGW}), which is valid for $\xi \gtrsim 1$. This leads to the following algebraic equation for $\xi$ as a function of $N \equiv \ln a$,
\begin{equation}
 \frac{{\rm e}^{4\pi\xi}}{\xi^6} = 
  2.3 \cdot 10^6 \, \frac{M_{\rm Pl}^2}{H^2}
  \left[
  \frac{\pi^2M_{\rm Pl}^2}{2H^2}
  P_{\rm GW}(H {\rm e}^N) - 1
  \right] \; . \label{eqn:EQforxi}
\end{equation} 
Since $H$ is a function of $N$, the right hand side of (\ref{eqn:EQforxi}) is an already known function of $N$.~\footnote{We stress that we are working in a regime in which the pseudo-scalar $\psi$ and the vector quanta give a negligible contribution to the background dynamics and to the scalar perturbations. Here we are working in the hypothesis that the scalar perturbations allow us to reconstruct the inflaton potential, and therefore $H \left( N \right)$, with the required accuracy.}  By solving the algebraic equation (\ref{eqn:EQforxi}) with respect to $\xi$, we thus obtain $\xi$ as a function of $N$, $\xi=\xi(N)$.

We now show that the axion potential $U(\psi)$ can be reconstructed from
$\xi(N)$ and $H(N)$. First, from the definition (\ref{eom-gauge}) and
(\ref{def-deltaxi}) of $\xi$ and $\delta_{\xi}$, we have 
\begin{eqnarray}
 \dot{\bar \psi} & = & 2fH\xi \; , 
 \label{eqn:dotpsi}\\
 \ddot{\bar \psi} & = & 2fH^2\xi\, (\delta_{\xi}-\epsilon) \; ,
  \label{eqn:ddotpsi}
\end{eqnarray} 
where $\delta_{\xi}=d\ln\xi/dN$ can be considered as a function of
$N$. By substituting (\ref{eqn:dotpsi}) and (\ref{eqn:ddotpsi}) to
(\ref{eqn:EOMpsi}), we obtain 
\begin{equation}
 U'(\bar \psi) = -2fH^2\xi\, (3+\delta_{\xi}-\epsilon) \; . 
 \label{eqn:U'}
\end{equation} 
By using (\ref{eqn:dotpsi}) and (\ref{eqn:U'}), we can express $d\bar\psi/dN$ and $dU/dN$ as functions of $N$.
By integrating them with respect to $N$, we obtain $\bar \psi$ and $U$ as functions of $N$, 
\begin{eqnarray}
 \bar \psi & = & \bar \psi(N) \equiv 2f \int \xi\, dN \; , \nonumber\\
 U & = & U(N) \equiv 
  -4f^2\int H^2\xi^2\, (3+\delta_{\xi}-\epsilon)\, dN \; .
  \label{eqn:psi-U}
\end{eqnarray}
Finally, the elimination of $N$ from $U=U(N)$ and $\bar \psi=\bar \psi(N)$ results in the reconstruction of the pseudo-scalar potential $U = U [ N ( \bar \psi )] =U(\bar \psi)$. 

Eqs.~(\ref{eqn:psi-U}) define the axion potential $U$ as a single-valued
regular function of the pseudo-scalar $\bar \psi$. This requires that
$H$, $d\ln H/dN$, $\xi$ and $d\ln\xi / dN$ are regular.
On the other hand, eq.~(\ref{eqn:EQforxi}) is valid for $\xi \gtrsim 1$, and in such cases,  eq.~(\ref{eqn:EQforxi}) can be inverted to give $\xi$ as a single-valued function of $N$. This then requires
\begin{equation}
 P_{\rm GW}(k) \geq 
 P_{\rm GW,0}(k) 
  \left[1 +  
   0.61 \, P_{\rm GW,0}(k) \right],
\end{equation}
where $P_{\rm GW,0}(k) \equiv \left. 2H^2 / (\pi^2M_{\rm Pl}^2)
\right|_{k=aH}$ is the contribution from vacuum fluctuations. Since the
second term in the square brackets is much smaller than unity, this
condition basically tells that the tensor spectrum must be larger than
the contribution from the vacuum fluctuations in order for $\xi(N)$ to
be reconstructable. Once $\xi(N)$ is reconstructed under this condition,
the regularity of $\xi$ and $d \ln \xi / d N$ is automatic, provided
that the tensor spectrum $P_{\rm GW}(k)$ is a smooth function of $k$ and
that $H$ is a smooth function of $N$. 

We have shown that the pseudo-scalar potential $U (\psi)$ can be
reconstructed, given the GW spectrum $P_{\rm GW}$ (and the scalar
spectrum $P_\zeta$, to find $H (N)$). This reconstruction in practice
relies on the experimental sensitivities to deviation from the
single-field relation $r = 16 \epsilon$. The right-hand side of
(\ref{eqn:EQforxi}) would vanish if this relation holds exactly (recall
that in our model the scalar spectrum has been shown to be standard by
(\ref{eqn:P-Pzeta}), i.e. $P_\zeta \simeq {\cal P}$), and $\xi(N)$
could not be obtained. Thus the reconstruction is possible only if the
relation $r = 16 \, \epsilon$ is found to be violated beyond the
experimental precision. 

\begin{figure}
\centering
\includegraphics[width=0.7\textwidth]{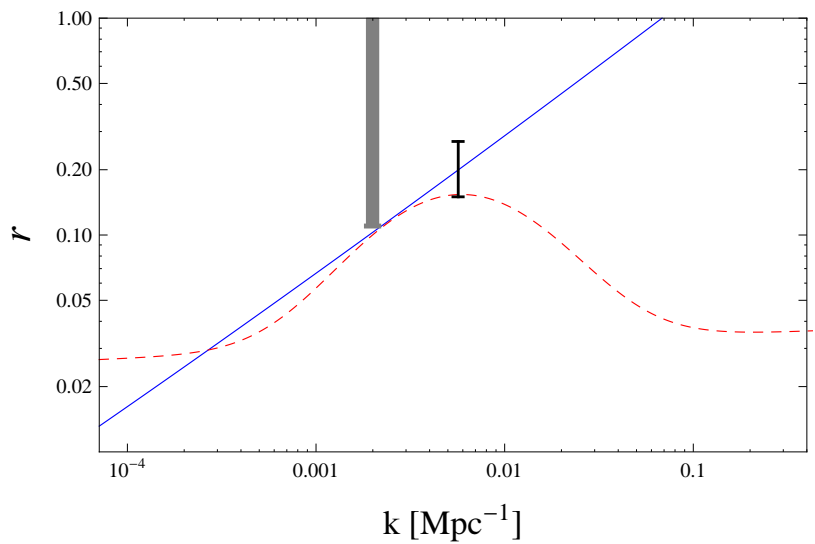}
\caption{Two illustrative spectra of the tensor-to-scalar ratio. The blue curve corresponds to the case in Example I, and the red dashed curve to the case in Example II. See the corresponding subsections for the details. The Planck upper bound $r < 0.11$ at $k = 0.002 \; {\rm Mpc}^{-1}$ and the BICEP2 measurement $r = 0.20_{-0.05}^{+0.07}$ at $k = 0.0057 \; {\rm Mpc}^{-1}$ ($\ell \approx 80$) are also shown.
}
\label{fig:r-vs-k}
\end{figure}

In the following subsections, we consider some instructive examples. For illustrative purposes,  a spectrum of the tensor-to-scalar ratio $r$ for each of the examples is shown in Figure \ref{fig:r-vs-k}. As seen from the figure, a blue spectrum between the Planck satellite and BICEP2 scales can easily be obtained in this mechanism, and the possible tension between the two experiments can be accounted for. Regarding the growth in the blue curve in FIG.~\ref{fig:r-vs-k}, illustrating Example I, we caution that we are assuming that the $\delta_\xi$ remains constant at the CMB scales, but we are not implying that the growth continues indefinitely at smaller scales.%
\footnote{
We assume that the growth of the blue solid curve in Figure \ref{fig:r-vs-k} is cut off for the modes of smaller scales. This can be done, for example, once the pseudo-scalar field $\psi$ starts oscillating at its potential minimum, which effectively makes the production inefficient.
}
We verified that these examples respect the non-gaussianity limit discussed in Section \ref{gaugeprod}.

\subsection{Example I: constant $\delta_{\xi}$ and continuously growing $r$}
\label{subsec:exampe1}

In this example, we consider constant $\epsilon(N)$ and $\delta_{\xi}(N)$, i.e.,
\begin{equation}
 \epsilon = \mbox{const.} \; , \quad \delta_{\xi} = \mbox{const.} \; ,
 \label{constant-delxi}
\end{equation}
leading to
\begin{equation}
 H(N) = H_c \, {\rm e}^{-\epsilon (N-N_c)} \; , \quad 
 \xi(N) = \xi_c \, {\rm e}^{\delta_{\xi}(N-N_c)} \; .
\end{equation} 
The corresponding tensor spectrum is
\begin{equation}
 P_{\rm GW}(k) = \frac{2H_c^2}{\pi^2M_{\rm Pl}^2}
  \left(\frac{k}{k_c}\right)^{-\frac{2\epsilon}{1-\epsilon}}
  \Bigg\{ 1 + 4.3 \cdot 10^{-7} \frac{H_c^2}{M_{\rm Pl}^2 \xi_c^6}
  	\left(\frac{k}{k_c}\right)^{-\frac{2(\epsilon+3\delta_{\xi})}{1-\epsilon}}
	\exp\left[4\pi\xi_c
	       \left(\frac{k}{k_c}\right)^{\frac{\delta_{\xi}}{1-\epsilon}}
	      \right]\Bigg\}, 
\end{equation}
where $k_c \equiv a_c H_c$.

In this case, from (\ref{eqn:psi-U}) we obtain
\begin{eqnarray}
 \bar \psi(N) & = & \psi_c +
  \Delta \psi_c \, e^{\delta_{\xi}(N-N_c)}, \nonumber\\
 U(N) & = &  U_c 
  + \Delta U_c
  \left[1-e^{2(\delta_{\xi}-\epsilon)(N-N_c)}\right] \; ,
\end{eqnarray} 
where $\psi_c$ and $U_c$ are integration constants, and
\begin{equation}
\Delta \psi_c \equiv \frac{2 \xi_c f}{\delta_\xi} \; , \quad \Delta U_c \equiv \frac{2\xi_c^2 f^2 H_c^2(3+\delta_{\xi}-\epsilon)}
  {\delta_{\xi}-\epsilon} \; .
\end{equation}
By eliminating $N$ we obtain
\begin{equation}
 U(\bar \psi) = U_c 
  + \Delta U_c
  \left[1-
   \left( \frac{\bar \psi-\psi_c}{\Delta \psi_c}
   \right)^{\frac{2(\delta_{\xi}-\epsilon)}{\delta_{\xi}}}
  \right] \; .
\end{equation}

\begin{figure}
 \begin{center}
  \includegraphics[width=0.7\textwidth]{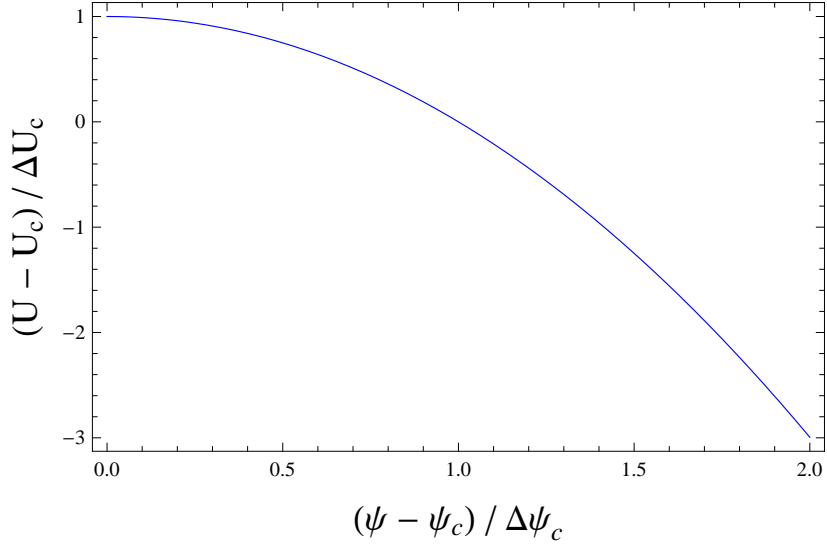}
  \caption{Reconstructed pseudo-scalar potential for constant $\epsilon = 10^{-5}$ and $\delta_\xi = 0.012$ in Example I, corresponding to the blue solid curve in Figure \ref{fig:r-vs-k}. The horizontal axis is 
  $(\bar \psi-\psi_c)/\Delta\psi_c$ and the vertical axis is 
  $(U-U_c)/\Delta U_c$.
  \label{fig:psipot-cons}}
  \end{center}
  \end{figure}

An illustrative shape of this reconstructed $U(\bar \psi)$ is shown in Figure \ref{fig:psipot-cons}, with $\epsilon = 10^{-5}$ and $\delta_\xi = 0.012$. The value of $n_T = 0.6$ at the domination of the sourced GWs is chosen to give $r = 0.2$ at the BICEP2 scale and to satisfy the upper bound from the Planck. This choice of parameters gives $\xi_c = 4.6$ and corresponds to the blue solid curve in Figure \ref{fig:r-vs-k}. The COBE normalization fixes $H_c = 1.4 \cdot 10^{-6} M_{\rm Pl}$, and the value of $k_c$ is fixed at the BICEP2 scale, $k_c = 0.0057 \; {\rm Mpc}^{-1}$.

\subsection{Example II: Gaussian $\xi(N)$ and bump in $r$}
\label{subsec:example2}

In the second example, we again consider $\epsilon(N)$ as a constant, while adopting the Gaussian ansatz for $\xi(N)$, that is,
\begin{equation}
 \epsilon = \mbox{const.} \; , \quad \xi(N) = \xi_c \, {\rm e}^{-\alpha^2(N-N_c)^2} \; , \label{eqn:Gaussian-xi}
\end{equation}
where $\xi_c$ ($>0$), $\alpha$ ($>0$) and $N_c$ are constants. This again leads to
\begin{equation}
 H(N) = H_c \, {\rm e}^{-\epsilon (N-N_c)} \; ,
\end{equation}
and now the tensor spectrum is
\begin{equation}
 P_{\rm GW}(k) 1= \frac{2H_c^2}{\pi^2M_{\rm Pl}^2}
  \left(\frac{k}{k_c}\right)^{-\frac{2\epsilon}{1-\epsilon}}
  \bigg[1 + 4.3 \cdot 10^{-7} \frac{H_c^2}{M_{\rm Pl}^2}
	  \left(\frac{k}{k_c}\right)^{-\frac{2\epsilon}{1-\epsilon}}
	  \frac{e^{4\pi\tilde{\xi}(k)}}
	  {\tilde{\xi}(k)^6}
	 \bigg] \; , 
\end{equation}
where $k_c\equiv a_c H_c$ and
\begin{equation}
\tilde{\xi}(k) =
  \xi_c\exp\left\{-\left[\frac{\alpha}{1-\epsilon}
		    \ln\left(\frac{k}{k_c}\right)\right]^2\right\} \; .
\end{equation} 
This tensor spectrum has a peak around
$k=k_c$.

In this case, from (\ref{eqn:psi-U}) we find
\begin{eqnarray}
 \bar \psi(N) & = & \psi_c + \Delta\psi_g \;
  {\rm erf} \left[ z(N) \right], \nonumber\\
 U(N) & = &  U_c + \Delta U_g
  \Bigg\{1 - {\rm e}^{-2z(N)\left[z(N)+ \frac{\epsilon}{\alpha} \right]}
  \nonumber\\
 & & \qquad\quad
  - \frac{3\sqrt{2\pi}}{2\alpha} \, {\rm e}^{\frac{\epsilon^2}{2\alpha^2}}
  \left[
   {\rm erf}\left(\sqrt{2} \, z(N) + \frac{\epsilon}{\sqrt{2} \alpha} \right)
   - {\rm erf}\left( \frac{\epsilon}{\sqrt{2} \alpha} \right)\right]
	\Bigg\} \; , 
\end{eqnarray} 
where $\psi_c$ and $U_c$ are integration constants, 
${\rm erf}(x)=\frac{2}{\sqrt{\pi}}\int_0^xe^{-t^2}dt$ is the error function, and
\begin{equation}
\Delta\psi_g \equiv \frac{\sqrt{\pi}\xi_cf}{\alpha} \; , \quad
\Delta U_g \equiv 2 \xi_c^2 f^2 H_c^2 \; , \quad
 z(N) = \alpha(N-N_c) \; .
\end{equation} 
Inverting the relation $\bar \psi = \bar \psi (N)$ to 
$N = {\rm erf}^{-1} \left( \frac{\bar \psi - \psi_c}{\Delta \psi_g} \right)$, 
and substituting it to $U(N)$, we can reconstruct $U (\bar \psi)$.  

\begin{figure}
 \begin{center}
  \includegraphics[width=0.7\textwidth]{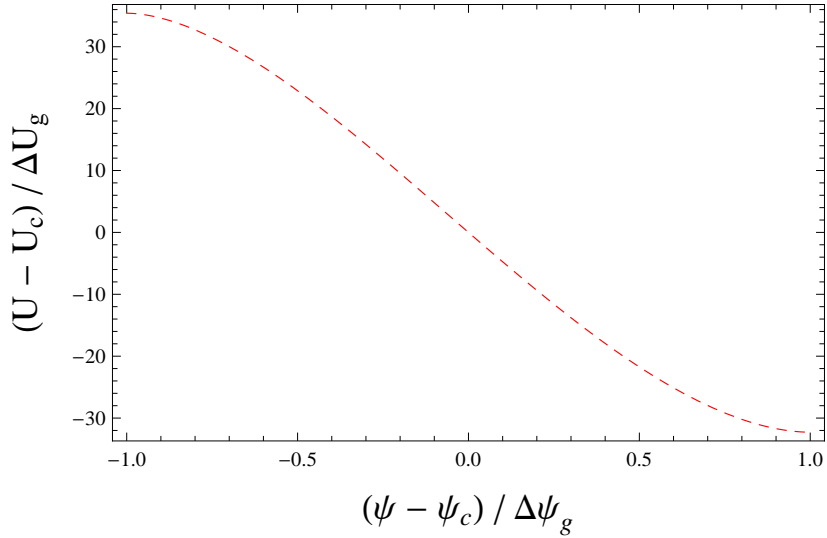}
  \caption{Reconstructed pseudo-scalar potential for the ansatz (\ref{eqn:Gaussian-xi}), corresponding to the red dashed curve in Figure \ref{fig:r-vs-k}.
  The parameters are chosen to be $\epsilon = 2.3 \cdot 10^{-3}$, $\xi_c = 3.6$, and $\alpha = 0.11$ for illustration.
  The horizontal axis is $(\bar \psi-\psi_c)/\Delta\psi_c$ and the vertical axis is $(U-U_c)/\Delta U_c$.
  \label{fig:psipot-gauss}}
  \end{center}
  \end{figure}

An illustrative shape of the reconstructed $U(\bar \psi)$ corresponding to the red
dashed curve in Figure \ref{fig:r-vs-k} is shown in
Figure \ref{fig:psipot-gauss}, with $\epsilon = 2.3 \cdot 10^{-3}$, $\xi_c = 3.6$ and $\alpha = 0.11$. The COBE normalization fixes $H_c = 2.1 \cdot 10^{-5} M_{\rm Pl}$, and the value of $k_c$ is again fixed at the BICEP2 scale, $k_c = 0.0057 \; {\rm Mpc}^{-1}$.

\section{Summary and discussions}
\label{sec:conclusions}

We have studied gravitational waves sourced by particles produced during
inflation. Our model consists of a pseudo-scalar field that is
subdominant during and after inflation, and a $U(1)$ gauge field in a
hidden sector. Gauge field particles are produced through an axial
coupling as the pseudo-scalar rolls its potential. The stress energy
tensor of produced particles then acts as a source term for the Einstein
equation and generates gravitational waves. We have shown that
gravitational waves generated in this way can acquire a blue spectrum, while the scalar spectrum is kept standard,
and can be responsible for reconciling the (possible) tension between the
recent BICEP2 result and the Planck constraints.

The standard relation $r=16\epsilon$ and the consistency relation
$n_T=-r/8$ hold for vacuum fluctuations in any single-field slow-roll inflationary
models. However, both relations do not necessarily hold in more general
setups. In the model studied in the present paper, as shown in
Figure \ref{fig:eps-xi}, the amount of deviation from the standard
relation is parameterized by the quantity $\xi$ that roughly measures
the change of the pseudo-scalar in one Hubble time, weighted with the
axial coupling constant. This is what makes it possible for the model
under consideration to generate a blue tensor spectrum (see eq.~(\ref{tensor-tilt})).

This mechanism can be combined with any inflationary models, provided
that an inflaton (or another field, such as a curvaton) can produce 
an observationally viable spectrum of scalar perturbations and that the
amplitude of tensor perturbations from vacuum fluctuations is not too 
large. This would help reconcile many low-scale inflationary models with
observational data if the large tensor-to-scalar ratio reported by the
BICEP2 is confirmed in the future. 

Our scenario can be discussed generally in a field-theoretical context though it also finds a natural home in string theory. Axion-like particles are ubiquitous in string compactifications as they arise from Kaluza-Klein reduction of various antisymmetric fields on cycles of the internal space.
Interestingly, the backreaction constraint of our scenario (see eq.~(\ref{backreaction})) requires the axion decay constant to satisfy $f/M_P \gtrsim 10^{-4}$ which  falls into the typical range one finds in string theory models, especially
those with Grand Unified Theories like phenomenology \cite{Choi:1985je,Banks:2003sx,Svrcek:2006yi}. The hidden nature of the axion and $U(1)$ gauge field can be ensured by imposing some topological constraints on the underlying string construction. For example, if the inflaton and the hidden $U(1)$ are realized on the worldvolume of different D-branes, these constraints amount to requiring that the axion does not serve as a portal  between the two sectors (unlike the
St\"uckelberg portal recently investigated in \cite{Feng:2014eja,Feng:2014cla}).
It would be interesting to find concrete string models realizing our scenario.

One of the robust predictions of the model is parity violation in tensor
perturbations. If the sourced gravitational waves are dominant over
those from vacuum fluctuations, the tensor spectrum is almost maximally
parity violating. It is thus expected that Planck (with upcoming B mode
polarization data) and Spider will be able to observe parity violation
in the sky at the $1\sigma$ level. Another important prediction is
non-Gaussianity in the tensor perturbations. While the predicted
non-Gaussianity is consistent with constraints to date, the Planck experiment (with
upcoming B mode polarization data) can start probing this mechanism 
by the non-Gaussianities in tensor perturbations. 

We have argued that gravitational waves sourced by particle production
can save some low-scale inflationary models that would otherwise be in
conflict with the high tensor-to-scalar ratio. Similarly, some early 
universe scenarios alternative to inflation appear to be in conflict
with the high tensor-to-scalar ratio and such a situation may also be 
ameliorated by particle production followed by generation of
gravitational waves. Detailed investigation in concrete setups seems
worthwhile as a future work.

\subsection*{Acknowledgments}

We thank  Kiwoon Choi and Lorenzo Sorbo for helpful discussions.
The work of SM and RN is supported by the World Premier International
Research Center Initiative (WPI Initiative), MEXT, Japan. The work of SM
is supported in part by Grant-in-Aid for Scientific Research 24540256,
21111006 and 21244033, MEXT, Japan. 
This work of MP is supported in part by DOE grant DE-FG02-94ER-40823 at the University of Minnesota.
The work of GS
is supported in part by  DOE grant DE-FG-02-95ER40896 at the University of Wisconsin.

\bibliographystyle{apsrmp}

\end{document}